\documentclass[aps, prl, twocolumn, groupedaddress, showpacs]{revtex4-1}

\usepackage{graphicx}
\usepackage{amsmath}
\usepackage{amssymb}
\usepackage{color}
\usepackage{dcolumn}
\usepackage{epsfig}
\usepackage{bm}

\begin{document}

\title{The Strain Derivatives of $T_c$ in HgBa$_2$CuO$_{4+\delta}$: CuO$_2$ Plane Alone Is Not Enough}

\author{Shibing Wang$^{1,2}$, Jianbo Zhang$^3$, Jinyuan Yan$^4$, Xiao-Jia Chen$^{5,6}$, Viktor Struzhkin$^5$, Wojciech Tabis$^7$, Neven Bari\v{s}i\`{c}$^{7,8}$, Mun Chan$^7$, Chelsey Dorow$^7$, Xudong Zhao$^{7,9}$, Martin Greven$^7$, Wendy L. Mao$^{1,10}$, Ted Geballe$^{11}$}

\address{$^1$Department of Geological and Environmental Sciences, Stanford University, Stanford, CA 94305, USA }
\email[]{Email: shibingw@stanford.edu}
\address{$^2$SIMES, SLAC National Accelerator Laboratory,
Menlo Park, CA 94025, USA}
\address{$^3$Department of Physics, South China University of Technology, Guangzhou 510640,China}
\address{$^4$Advanced Light Source, Lawrence Berkeley National Laboratory, Berkeley, CA 94720, and Earth and Planetary Sciences, University of California, Santa Cruz, CA 95064, USA }
\address{$^5$Geophysical Laboratory, Carnegie Institution of Washington, Washington, DC 20015, USA}
\address{$^6$Center for High Pressure Science and Technology Advanced Research, Shanghai 201203, China}
\address{$^7$School of Physics and Astronomy, University of Minnesota, MN, USA}
\address{$^8$Institute of Physics, HR-10000 Zagreb, Croatia}
\address{$^9$State Key Lab of Inorganic Synthesis and Preparative Chemistry, Jilin University, Changchun 130012, China}
\address{$^{10}$Photon Science, SLAC National Accelerator Laboratory, Menlo Park, CA 94025, USA}
\address{$^{11}$Department of Applied Physics,and Geballe Laboratory for Advanced Materials, Stanford University, Stanford, CA 94305, USA}

\date{\today}

\begin{abstract}

The strain derivatives of $T_c$ along the $a$ and $c$ axes have been determined for HgBa$_2$CuO$_{4+\delta}$ (Hg1201), the simplest monolayer cuprate with the highest $T_c$ of all monolayer cuprates ($T_c$ = 97 K at optimal doping). The underdoped compound with the initial $T_c$ of 65~K has been studied as a function of pressure up to 20~GPa by magnetic susceptibility and X-ray diffraction (XRD). The observed linear increase in $T_c$ with pressure is the same as previously been found for the optimally-doped compound.  The above results have enabled the investigation of the origins of the significantly different $T_c$  values of optimally doped Hg1201 and the well-studied compound La$_{2-x}$Sr$_{x}$CuO$_{4}$ (LSCO), the latter value of $T_c$ = 40~K being only about 40\% of the former. Hg1201 can have almost identical CuO$_6$ octahedra as LSCO if specifically strained. When the apical and in-plane CuO$_2$ distances are the same for the two compounds, a large discrepancy in their $T_c$ remains. Differences in crystal structures and interactions involving the Hg-O charge reservoir layers of Hg1201 may be responsible for the different $T_c$ values exhibited by the two compounds.

\end{abstract}


\maketitle

More than two decades after the discovery of high temperature superconductors with superconducting transition temperature ($T_c$) above the liquid nitrogen boiling point, the mechanisms leading to such extraordinary high $T_c$ values remain unclear. Correlated electrons within the copper-oxygen planes form Cooper pairs. $T_c$ is a function of cation or oxygen doping. It rises to a maximum at optimal doping and then falls in a "dome" like trajectory~\cite{Yamamoto,Liang}. When subject to pressure, $T_c$ of some optimally doped compounds increases at a rate of 1-2~K/GPa before saturating at a certain pressure. Among these cuprates is the mercury family, which are model systems with copper-oxygen planes sandwiched by mercury oxygen planes: HgBa$_2$Ca$_{n-1}$Cu$_n$O$_{2n+2+\delta}$ ($n$=1,2,3, ...9)\cite{Gao_1994,Iyo_9layer}. The trilayer compound (n=3) holds the record $T_c$ of 164 K when compressed to 30~GPa ~\cite{Gao_1994}.

Strain effects on the $T_c$ of the cuprate superconductors
provide important information to help guide the development of adequate theoretical models and, potentially, for the design of materials with higher values of $T_c$. There have been a number of high pressure studies on optimally doped Hg1201, investigating how lattice parameters, atomic positions, and $T_c$ changes under both hydrostatic and uniaxial pressure~\cite{Gao_1994, Hunter_1994, Eggert_1994, Hardy_PRL_2010}. The uniaxial d$T_c$/d$P_l$ ($l=a,b,c$) has been found from the Ehrenfest relationship \(dT_c/dP_l=\Delta\alpha_lV_mT_c/\Delta C_p\) using experimental values of the thermal expansion ($\alpha_l$) and heat capacity ($\Delta C_p$)~\cite{Schilling_chapter}.  The hydrostatic d$T_c$/d$P$, on the other hand, is directly determined from either susceptibility or transport measurements. These values are essentially the
\emph{stress} derivatives of $T_c$. To test current theories, the \emph{strain} coefficients d$T_c$/(d$l/l$) are particularly useful.
By obtaining the strain derivatives of $T_c$ along the different crystallographic axes, we aim to establish that the large discrepancy in $T_c$ between Hg1201 and LSCO cannot be explained by interactions confined to the CuO$_2$ planes alone.

In this letter, we present the dependence of $T_c$ and structure on pressure for underdoped single crystals of Hg1201 with an ambient $T_c$ at 65~K measured up to 20~GPa in diamond anvil cells (DACs). We find that the rate of $T_c$ increase agrees with that of optimally doped Hg1201 ~\cite{Gao_1994, Hunter_1994, Klehe_1993} for a wide pressure range. The effect of pressure, either uniaxial or hydrostatic, on $T_c$ is linear, $i.e.$ d$T_c$/d$P_l$ and d$T_c$/d$P$ (hydrostatic) are constant, up to 10 GPa for both underdoped and optimally doped Hg1201, which suggests that pressure is tuning interactions that are independent of the carrier density \cite{chen0}.

\begin{figure}[tbp]
\vspace{-8mm}
\includegraphics[width=0.9\columnwidth]{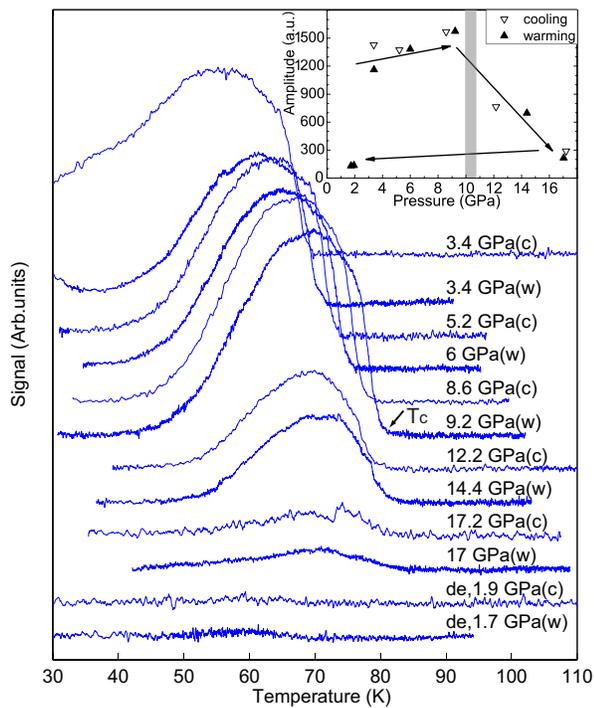}\\
\caption{(Color online)In-phase component of susceptibility signal measured during both cooling and warming cycles at each pressure run. The run started with 3.4~GPa and was increased to 17~GPa. Pressure was then released to 1.9~GPa immediately.'de' is short for decompression. Inset: Strength of the susceptibility signal as a function of pressure. Arrows indicate the measurement sequence. Gray bar indicates the pressure where sample starts to degrade.}\label{fig:1}
\end{figure}

The samples measured in the present experiment were grown with an encapsulation method and subsequently annealed to yield a $T_c$ of 65~K~\cite{Zhao_AdvMat_2006,Barisic_2008}. For the $T_c$ measurement, a 120$\times$80$\times$30~$\mu$m$^3$ single crystal was loaded into a Mao-Bell DAC made from hardened Be-Cu alloy. A nonmagnetic Ni-Cr alloy gasket pre-indented to 35~$\mu$m thick with a 250~$\mu$m diameter hole served as the sample chamber. Daphne 7373 was loaded into the gasket hole as a pressure medium. An AC circuit consisted of a signal coil around the diamonds, a compensating coil nearby and a larger pick up coil was used to measure susceptibility, detailed previously~\cite{Viktor_2002,chen1,chen2}. The synchrotron XRD experiment was conducted at Beamline 12.2.2 of Advanced Light Source (ALS) with incident x-ray wavelength of 0.6199~\AA. A sample from the same mother crystal was ground into a powder in an agate motar and was loaded to a symmetric DAC with a stainless steel gasket in a hole with 150~$\mu$m diameter; the diamond culet was 300~$\mu$m. Ne gas was loaded into the sample chamber as the pressure medium~\cite{GSECars_Gasloading}. Rietveld refinement was performed on the powder diffraction pattern. In both measurements, small ruby chips placed in the DACs were used for pressure calibration~\cite{Ruby_1986_JGR}.

\begin{figure}[tbp]
  \includegraphics[width=0.88\columnwidth]{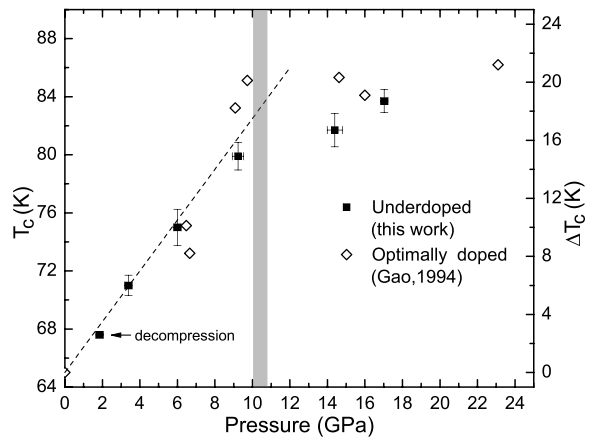}\\
  \caption{($T_c$ and $\Delta T_c$ vs pressure. Filled squares: $T_c$ of the underdoped sample measured in the warming cycle. Open diamonds: $\Delta T_c$ of optimally doped sample~\cite{Gao_1994}. The dashed line corresponds to d$T_c$/d$P$=1.75~K/GPa ~\cite{Klehe_1993}. Gray bar indicates the pressure where sample starts to degrade.}\label{fig:1b}
\end{figure}

Fig.~\ref{fig:1} shows the in-phase component of the modulated signal versus temperature for underdoped Hg1201. For each pressure run, the signal was measured during both cooling and warming cycles. $T_c$ is taken as the intersection of the extrapolated linear rise with the base line~\cite{Viktor_2002}. Pressures were measured 10-15~K above the transition temperature. When the sample was warmed up to 120~K, pressure was increased, and after 30~min of relaxation, $T_c$ was measured at the new pressure. The $T_c$ of underdoped Hg1201 increased from 65~K at ambient pressure to 84~K at 17~GPa. Upon reducing the pressure back to ambient~\cite{quench}, the high $T_c$ (84~K) was not retained, and the signal amplitude was not recovered.

The inset of Fig.~\ref{fig:1} shows that the amplitude of the signal increases with increasing pressure before decreasing significantly at 12~GPa. Previous resistivity measurements on optimally doped Hg1201 suggest that defects are introduced at high quasi-hydrostatic pressure causing irreversible degradation of the sample above 10 GPa~\cite{Gao_1994}.

Fig.~\ref{fig:1b} shows that $T_c$ increases linearly with applied pressure up to $\sim$10~GPa.  The increase of $T_c$ compared to ambient pressure ($\Delta T_c$) is also plotted to compare with the $\Delta T_c$ of optimally doped Hg1201 measured resistively~\cite{Gao_1994}. Two observations can be made: First, the linearity range of d$T_c$/d$P$ extends up to $\sim$10~GPa in Hg1201, approximately the same pressure above which the suceptibility measurements indicates sample degradation (Fig. 1); Second, the $\Delta T_c$  response of Hg1201 to pressure is almost identical for underdoped and optimally doped samples. Such an agreement of underdoped and optimally doped Hg1201 was previously observed only up to 1.7~GPa~\cite{Cao_1995}.

\begin{figure}[tbp]
  \vspace{0.3cm}
  \includegraphics[width=0.49\textwidth]{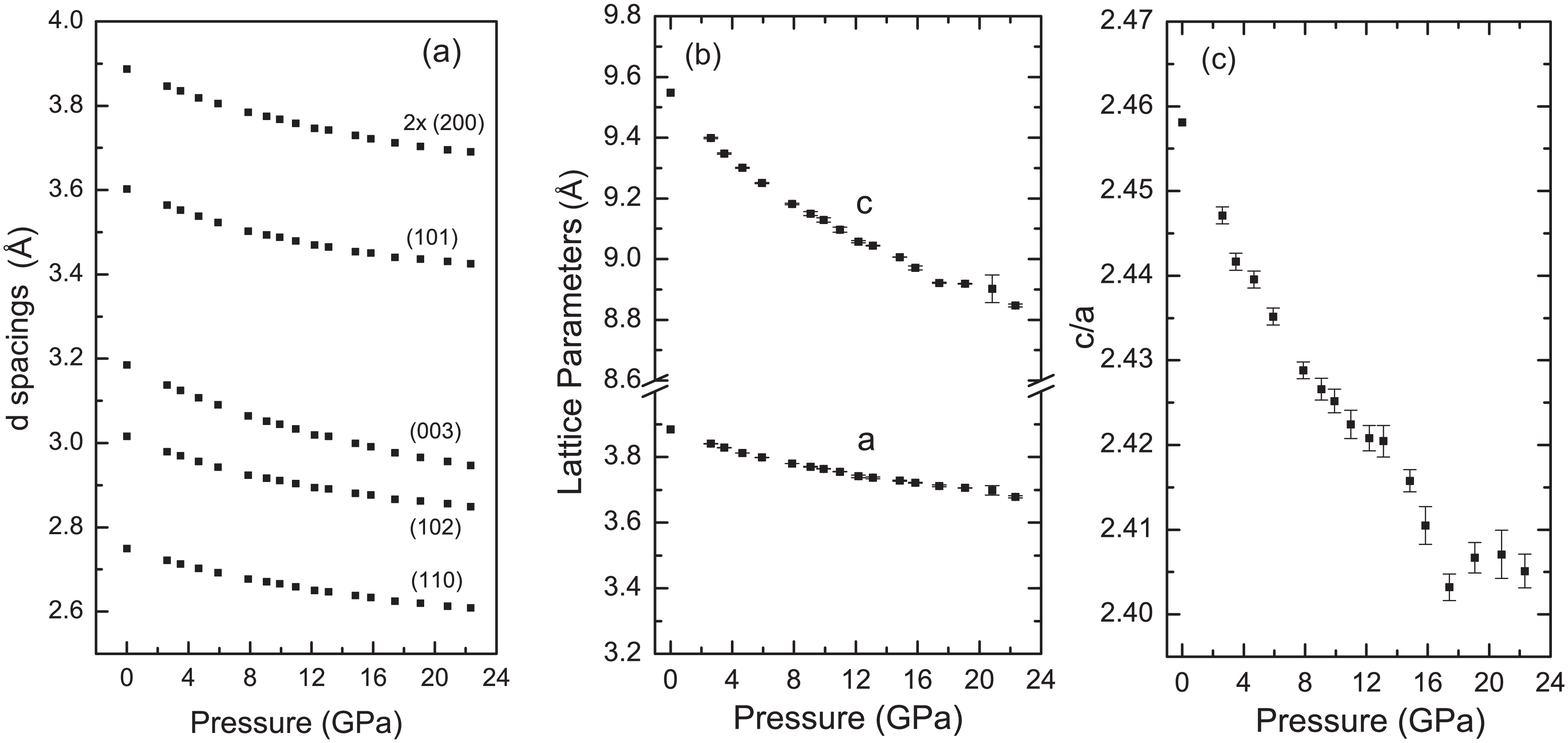}\\
  \caption{(a) The $d$-spacings for the (110), (102), (003), (101) and (200) Bragg reflections as a function of pressure for underdoped Hg1201. (b)Lattice parameters and (c) $c/a$ ratio as a function of pressure. }\label{fig:3}
\end{figure}

Structural information for Hg1201 is summarized in Fig.~\ref{fig:3}. The pressure dependence of the (003),(110) and (200) Bragg peak positions  indicates that lattice parameter $c$ decreases at a faster rate than $a$, consistent with a previous report for optimally doped Hg1201~\cite{Eggert_1994}. The lattice parameters and volume were fit to a third-order Birch-Murnaghan equation with $K_0$'=4~\cite{EOS}. We obtain axes and volume bulk moduli $Ka_0$, $Kc_0$, and $K_{V0}$ to be 83.6, 54.3, and 69.1~GPa respectively; the first two correspond to the $a$ and $c$ axial compressibilities $\kappa_a$, $\kappa_c$ ($\kappa_{a,c}$=1/(3$Ka_0$,c$_0$)) of $3.99 \times 10^{-3}$ and $6.13 \times 10^{-3}$ GPa$^{-1}$ at ambient pressure. These values agree well with those for optimal doping ~\cite{Eggert_1994, Hunter_1994, Balagurov_1999}, indicating that to first order, we can use these structure and elastic constants for Hg1201 for both the underdoped and optimally doped cases. Compressibilities at 7 and 11 GPa are given in Tab.~\ref{tab:2}. Due to peak broadening and weaker signals the refinement at higher pressure is less accurate. The $c/a$ ratio decreases approximately linearly up to $\sim$10 GPa, and exhibits a more complicated dependence at higher pressures (Fig.~\ref{fig:3}c). The anomalous region coincides with where the susceptibility signal decreases significantly (Fig.~\ref{fig:1}), and reflects the intrinsic sample change above 10-12~GPa. The identical $T_c$ responses to external pressure and similar $a$ and $c$ compressibilities for underdoped and optimally doped Hg1201 suggest that the rate at which the charge reservoir layer is brought toward the CuO$_2$ plane correlates with the rate of $T_c$ increase regardless of the initial charge carrier density.

We now focus on the strain derivative d$T_c$/(d$l/l$) for Hg1201. A series of uniaxial pressure and hydrostatic pressure experiments have been previously conducted on several cuprates, e.g. YBa$_2$Cu$_3$O$_{7-\delta}$, Tl$_2$Ba$_2$CuO$_{6+\delta}$,  Hg1201~\cite{Hardy_PRL_2010, Sadewasser_Tl-Y-Hg_2000, Sadewasser_Hg1201_1999,chen4}.  d$T_c$/d$P_l$ ($l$=$a$, $b$, or $c$) were obtained from the Ehrenfest relation. This is thermodynamically accurate  for  mean-field transitions, but it introduces some uncertainty in the Hg1201 case, where the $C_p$ anomaly spreads over two decades in temperature with no obvious discontinuous jump~\cite{vanHeumen_PRB_2007}.  With the compressibilities of $a$ and $c$ from our hydrostatic pressure XRD experiment, and making the reasonable assumption that Poisson's ratio  \(-\frac{dc/c}{da/a}=-\frac{db/b}{da/a}=0.2\)~\cite{Poisson}, we can obtain the relevant terms in the strain-stress compliance matrix of a tetragonal system (see Supplemental Materials for details). We use the widely accepted (and verified in the present work) value d$T_c$/d$P$=1.75~K/GPa~\cite{Klehe_1993,Schilling_chapter} and the best available d$T_c$/d$P_a$ = 2.3~K/GPa or d$T_c$/d$P_c$ = -3.6~K/GPa from uniaxial pressure experiment~\cite{Hardy_PRL_2010}. The calculated values of d$T_c$/(d$c/c$) and d$T_c$/(da/a) at different pressure are shown in Tab.\ref{tab:2}. Even though d$T_c$/d$P_c$ is larger in magnitude than d$T_c$/d$P_a$, the actual $T_c$ response to the $c$-axis strain is smaller. The ratio of the magnitude of d$T_c$/d$a$ - to - d$T_c$/d$c$ lies between 3.8-4.5, and d$T_c$/(d$a/a$) - to - d$T_c$/(d$c/c$) is 1.5-1.8 in Hg1201 at ambient pressure.

For  uniaxial pressure along the $c$-axis, the compression is accompanied by the expansion of the other two axes, i.e. \(dT_c/dP_c = \frac{\partial T_c}{\partial c}\frac{\partial c}{\partial P_c}+2\frac{\partial T_c}{\partial a}\frac{\partial a}{\partial P_c}\): both terms are negative with applied uniaxial pressure $P_c$. The large negative value of d$T_c$/d$P_c$ is  from the combination of $c$-axis compression and $ab$ plane expansion. The $T_c$ derivatives of the strain, on the other hand, separate these effects, and give direct information on how $T_c$ changes with different axis independently.

\begin{table}[b]
  \caption{Geometry of the CuO$_6$ octahedra for Hg1201 and LSCO
  at different pressure and temperature conditions, and strain derivatives of $T_c$ for Hg1201. Lattice parameters, compressibilities are from this study. Values of Cu-O$_{\text{apical}}$ are extrapolated from neutron scattering study\cite{Hunter_1994}. $T_c$ for optimally doped Hg1201 is from\cite{Gao_1994}, its buckling angle is extrapolated from~\cite{Hunter_1994}. Structure of LSCO is from~\cite{Cava_La214_1987}, its $T_c$ is from~\cite{La214_Geballe}. The uncertainty of the strain derivatives of $T_c$ comes from the slight disagreement of the uniaxial and hydrostatic stress derivatives and the choice of Poisson's ratio.}\label{tab:2}

  \centering
  \begin{ruledtabular}
\begin{tabular}{ccccc}
   & Hg1201 & Hg1201 & Hg1201 & La$_{1.85}$Sr$_{0.15}$CuO$_4$  \\

  Condition & ambient & 7~GPa & 11~GPa & 60~K  \\
  \hline
  a(\AA) & 3.885 & 3.78 & 3.754 & 3.78 \\
  c(\AA) & 9.549 & 9.205 & 9.089 & 6.59 \\
  Cu-O$_{\text{apical}}$ (\AA) & 2.789 & 2.552 & 2.417 & 2.41 \\
  Buckling (deg) & 180 & 180 & 180 & 175.5 \\
  $T_c$ (K) & 97 & 108 & 116 & 40 \\
  $\kappa_a$ (10$^{-3}$/GPa) & 3.99& 3.01 & 2.66 & \\
  $\kappa_c$ (10$^{-3}$/GPa) & 6.14& 4.11 & 3.49 & \\
  d$T_c$/(d$a/a$)(K) & -433(50) & -565(60) & -638(70) \\
  d$T_c$/(d$c/c$)(K) & 278(60) & 402(80) & 469(100) & \\

\end{tabular}
\end{ruledtabular}
\end{table}

Our calculation of d$T_c$/(d$l/l$) for Hg1201 provides the means for comparing the $T_c$ values of different families of cuprate superconductors. Here we compare the single-layer optimally-doped LSCO ($T_c$=40~K) with Hg1201 ($T_c$=97~K). With hydrostatic pressure, $T_{c,max}$ of LSCO reaches 42~K at 4~GPa, whereas for Hg1201 it reaches 118~K at 23~GPa. Hg1201 and LSCO differ in a number of ways, specifically: LSCO has a body centered structure and transforms to orthorhombic at low temperature which buckles the CuO$_2$ planes~\cite{Cava_La214_1987}, while Hg1201 has a simple tetragonal structure; the former has a shorter interlayer distance and apical oxygen distance and smaller CuO$_2$ plane area; in addition, differences in disorder have been noted~\cite{Eisaki_2004}. We aim to discern what are the contributing factors in the following discussion.

The lattice parameters and sizes of the CuO$_6$ octahedra of Hg1201 at different pressures are shown in Tab.~\ref{tab:2}: at 7 GPa the $ab$ plane of Hg1201 is of the same size as that of LSCO, while the apical oxygen distance is still 0.14~\AA ~larger than that of the latter. With d$T_c$/(d$c/c$) = 402~K(at P=7~GPa), $T_c$ is only reduced to 86~K, far above the $T_{c,max}$ of optimally doped LSCO (40~K)~\cite{La214_Geballe,Cava_La214_1987}. If we further increase pressure to 11~GPa, the apical oxygen distance of Hg1201 matches that of LSCO. Then, expanding $a$ by 0.026~\AA ~from 3.754~\AA ~to 3.78~\AA(Tab.~\ref{tab:2}) for Hg1201 will only reduce $T_c$ by 4~K. While we are aware of the complexity of the Cu-O-Cu buckling angle of Hg1201~\cite{buckling}, the difference in buckling angle between Hg1201 and LSCO would not account for much: High pressure reduces the buckling angle of LSCO to nearly 180$^{\circ}$ and makes the structure tetragonal~\cite{Yamada_1992} but only increases its $T_c$ for a few Kelvin~\cite{Mori_1991}. A-site (La site) disorder in LSCO influences $T_c$ through the hybridization between the orbitals of the apical O(2$p_z$) and Cu(3$d_{r^2-3z^2}$)~\cite{Eisaki_2004}. However, for the oxygen doped La$_2$CuO$_{4+\delta}$, where A-site disorder does not exist and additional oxygen is confined to interstitial sites~\cite{Birgeneau}, its $T_c$ only rises to 42~K~\cite{Schirber_La214P_1993}.

After adjusting the geometrical difference in the CuO$_6$ octahedra of Hg1201 and LSCO, there still remains a 44 K difference in $T_c$ values between the two cuprates. A recent theoretical model which explicitly includes the Cu $d_{x^2-y^2}$, $d_{z^2}$ and 4$s$ orbitals qualitatively predicts correctly the larger $T_c$ value of Hg1201~\cite{Sakakibara_PRL_2010} and the sign of  d$T_c$/d$P_l$ and d$T_c$/d$P$~\cite{Sakakibara_Multiorbital_2012}. The model attributes the low $T_c$of LSCO to the compound's body-centered tetragonal structure, in the close proximity of apical oxygen atoms of neighboring CuO$_2$ layers causes an elevation of the $d_{z^2}$ Wannier orbital~\cite{Sakakibara_Wannier_2012}.

However, the effect of the Hg-O layers seems to be more than merely separating the CuO$_6$ octahedra, as they exhibit a high degree of polarizability and hence serve to screen long-range Coulomb interactions in the quintessential CuO$_2$ sheets~\cite{Sushkov_2011, Raghu_2012}. We note that the above considerations have focused on average bond distances and bond angles. There exists ample evidence from local bulk probes that the cuprates exhibit significant compound-specific local deviations from the average crystal structure~\cite{Egami_1993,Agrestini}, and that the charge distributions in both LSCO~\cite{Singer} and Hg1201~\cite{Rybicki} vary on the nanoscale. Based on modeling the disorder in the interstitial layers, it was concluded that the hole mean free path and the screening of the Coulomb repulsion in Hg1201 are substantially larger than in LSCO, hence contributing to the higher $T_c$~\cite{Sushkov_2011}. In order to fully account for the differences between the two compounds, further consideration of the screening of electronic inhomogeneity inherent to the CuO$_2$ planes may be necessary. In this context, it is important to note that the Hg-O layers in Hg1201 may have metallic character that could be enhanced at elevated pressure~\cite{Jorgensen_review, Illas_JCP}.

In summary, through high pressure susceptibility and structure measurement of underdoped Hg1201, we obtained the hydrostatic d$T_c$/dP and relevant elastic constants of the compound. Together with previously reported d$T_c$/d$P_l$, we have determined d$T_c$/(d$l/l$) for Hg1201. Our results show that $T_c$ is more sensitive to the strain change along the $a$-axis than $c$-axis. A comparison of strained Hg1201 to optimally doped LSCO indicates that to account for the large $T_c$ discrepancy theories need to consider factors beyond the geometry of the CuO$_6$ octahedra.

The authors are grateful for discussion with Profs. W. Nix, S. Raghu, D. Scalapino and Dr. G. Yu. The authors thank Dr. S. Tkachev for help with gas loading at Advanced Photon Source. SW, ZJB, XJC, VS, and WLM are supported by EFree, an Energy Frontier Research Center funded by the U.S. Department of Energy (DOE), Office of Science, Office of Basic Energy Sciences(BES) under DE-SG0001057. Travel to facilities is supported by Stanford Institute for Materials and Energy Science (DE-AC02-76SF00515). The work at the University of Minnesota was supported by DOE-BES under DE-SC0006858. ALS is supported by DOE-BES under DE-AC02-05CH11231.

\clearpage

\section{Appendix}

\subsection{I. Constructing the Strain-Stress Compliance matrix}
Hydrostatic high pressure experiments fix the stress, and one measures the strain through x-ray diffraction (XRD). Therefore, the compliance matrix shall be used. To start, we have
\[\epsilon_i=S_{ij}\sigma_i \]
where we choose the crystal coordinates  $\epsilon_1 = da/a$, $\epsilon_2 = db/b$, and $\epsilon_3=dc/c$. For a tetragonal crystal system S$_{i,j}$ is reduced to
\[
\left(
  \begin{array}{c}
    \epsilon_1 \\
    \epsilon_2 \\
    \epsilon_3 \\
    \epsilon_4 \\
    \epsilon_5 \\
    \epsilon_6 \\
  \end{array}
\right)
=
\left(
  \begin{array}{cccccc}
    s_{11} & s_{12} & s_{13} &  &  & s_{16} \\
    s_{12} & s_{11} & s_{13} &  &  & -s_{16} \\
    s_{13} & s_{13} & s_{33} &  &   &   \\
      &   &   & s_{44} &   &   \\
      &   &   &   & s_{44} &   \\
    s_{16} & -s_{16} &   &   &   & s_{66} \\
  \end{array}
\right)
\left(
  \begin{array}{c}
    \sigma_1 \\
    \sigma_2 \\
    \sigma_3 \\
    \sigma_4 \\
    \sigma_5 \\
    \sigma_6 \\
  \end{array}
\right)
\]

In hydrostatic compression with external pressure P, this becomes
\[
\left(
  \begin{array}{c}
    \epsilon_1 \\
    \epsilon_2 \\
    \epsilon_3 \\
  \end{array}
\right)
=
\left(
  \begin{array}{ccc}
    s_{11} & s_{12} & s_{13}  \\
    s_{12} & s_{11} & s_{13}  \\
    s_{13} & s_{13} & s_{33}  \\
  \end{array}
\right)
\left(
  \begin{array}{c}
    -P \\
    -P \\
    -P \\
  \end{array}
\right)
\]
which gives
\begin{equation}\label{1}
    \epsilon_1=\epsilon_2=-P(s_{11}+s_{12}+s_{13})
\end{equation}

\begin{equation}\label{2}
    \epsilon_3=-P(2s_{13}+s_{33})
\end{equation}

With high pressure XRD, the compressibilities $\kappa_a = - \epsilon_1/P$,$\kappa_c = - \epsilon_3/P$ are known.

In $c$-axis uniaxial loading with $P_c$,we have

\[
\left(
  \begin{array}{c}
    \epsilon_1 \\
    \epsilon_2 \\
    \epsilon_3 \\
  \end{array}
\right)
=
\left(
  \begin{array}{ccc}
    s_{11} & s_{12} & s_{13}  \\
    s_{12} & s_{11} & s_{13}  \\
    s_{13} & s_{13} & s_{33}  \\
  \end{array}
\right)
\left(
  \begin{array}{c}
    0 \\
    0 \\
    -Pc \\
  \end{array}
\right)
\]
which gives \(\epsilon_1=-s_{13}P_c\), \(\epsilon_3=-s_{33}P_c\)
and Poisson ratio
\(\nu_{13}\equiv -\dfrac{\epsilon_1}{\epsilon_3}=-\dfrac{s_{13}}{s_{33}}\).

In $a$-axis uniaxial loading with $P_a$, we have
\[
\left(
  \begin{array}{c}
    \epsilon_1 \\
    \epsilon_2 \\
    \epsilon_3 \\
  \end{array}
\right)
=
\left(
  \begin{array}{ccc}
    s_{11} & s_{12} & s_{13}  \\
    s_{12} & s_{11} & s_{13}  \\
    s_{13} & s_{13} & s_{33}  \\
  \end{array}
\right)
\left(
  \begin{array}{c}
    -P_a \\
    0 \\
    0 \\
  \end{array}
\right)
\]
which gives \(\epsilon_1=-s_{11}Pa\),\(\epsilon_2=-s_{12}Pa\),
\(\epsilon_3=-s_{13}Pa\)
and two poisson ratios
\(\nu_{31}\equiv -\dfrac{\epsilon_3}{\epsilon_1}=-\dfrac{s_{13}}{s_{11}}\),
\(\nu_{21}\equiv -\dfrac{\epsilon_2}{\epsilon_1}=-\dfrac{s_{12}}{s_{11}}\).

Since we do not have elastic data from uniaxial compression, we have to make reasonable assumptions here. The first attempt is to assume the value for the Poisson ratio. Specifically for Hg1201 which does not have a huge $a/c$ anisotropy, we assume $\nu_{31},\nu_{21}$ to be 0.2, a reasonable value for ceramics. Therefore,

\begin{equation}\label{3}
    \nu_{31}=-\frac{s_{13}}{s_{11}}=0.2
\end{equation}

\begin{equation}\label{4}
    \nu_{21}=-\frac{s_{12}}{s_{11}}=0.2
\end{equation}

With four unknowns $s_{11},s_{12},s_{13},s_{33}$, and four equations \eqref{1},\eqref{2},\eqref{3},\eqref{4} we get
\[s_{11} = \dfrac{\kappa_a}{1+\nu_{21}+\nu_{31}}\]
\[s_{12} = \dfrac{\nu_{21}\kappa_a}{1+\nu_{21}+\nu_{31}}\]
\[s_{13} = \dfrac{\nu_{31}\kappa_a}{1+\nu_{21}+\nu_{31}}\]
\[s_{33} = \kappa_c-\dfrac{2\nu_{31}\kappa_a}{1+\nu_{21}+\nu_{31}}\]

\subsection{II. Converting $dT_c/d\sigma$ to $dT_c/d\epsilon$}

After the analysis of the previous section, we can express d$T_c$/d$P_a$, d$T_c$/d$P_c$, and d$T_c$/d$P$ in d$T_c$/d$\epsilon_1$,d$T_c$/d$\epsilon_3$, by writing out the full derivatives of $T_c$:

\small

\[   \frac{dT_c}{dP_a} =\frac{\partial T_c}{\partial \epsilon_1}\frac{\partial \epsilon_1}{\partial P_a}+\frac{\partial T_c}{\partial \epsilon_2}\frac{\partial \epsilon_2}{\partial P_a}+\frac{\partial T_c}{\partial \epsilon_3}\frac{\partial \epsilon_3}{\partial P_a}
     =(s_{11}+ s_{12}) \frac{dT_c}{d\epsilon_1}+s_{13} \frac{dT_c}{d\epsilon_3}\]

 \[   \frac{dT_c}{dP_c} =2\frac{\partial T_c}{\partial \epsilon_1}\frac{\partial \epsilon_1}{\partial P_c}+\frac{\partial T_c}{\partial \epsilon_3}\frac{\partial \epsilon_3}{\partial P_c}
     =2 s_{13} \frac{dT_c}{d\epsilon_1}+s_{33} \frac{dT_c}{d\epsilon_3}\]


 \[   \frac{dT_c}{dP} =2\frac{\partial T_c}{\partial \epsilon_1}\frac{\partial \epsilon_1}{\partial P}+\frac{\partial T_c}{\partial \epsilon_3}\frac{\partial \epsilon_3}{\partial P}
     =2 (s_{11}+s_{12}+s_{13}) \frac{dT_c}{d\epsilon_1}+(2s_{13}+s_{33}) \frac{dT_c}{d\epsilon_3}\]

\normalsize

 The above three equations are not independent, abiding to the relationship d$T_c$/d$P$ = 2d$T_c$/d$P_a$+d$T_c$/d$P_c$.

     If we use the value of d$T_c$/d$P_a$ and d$T_c$/d$P$ from experiments and $s_{11}, s_{12}, s_{13}, s_{33}$ from the above section, we'll be able to solve the following linear equations
     \small
     \[
     \left(
       \begin{array}{cc}
         s_{12}+s_{13} & s_{13}   \\
         2(s_{11}+s_{12}+s_{13}) & 2s_{13}+s_{33}  \\
       \end{array}
     \right)
     \left(
       \begin{array}{c}
         dT_c/d\epsilon_1 \\
         dT_c/d\epsilon_3 \\
       \end{array}
     \right)
= \left(
    \begin{array}{c}
       dT_c/dP_a\\
       dT_c/dP\\
    \end{array}
  \right)
\]
\normalsize
and obtain the values for
\[\frac{dT_c}{d\epsilon_1}\equiv\frac{dT_c}{da/a} = a\frac{dT_c}{da}\]
\[\frac{dT_c}{d\epsilon_3}\equiv\frac{dT_c}{dc/c} = c\frac{dT_c}{dc}.\]

\subsection{III. Case study for HgBa$_2$CuO$_4$}

From experiments,we use the following parameters at ambient pressure,
\[dT_c/dP_c = -3.6 \text{K/GPa}\]
\[dT_c/dP = 1.75 \text{K/GPa}\]
\[\kappa_a = 3.99 \times 10^{-3}\text{/GPa}\]
\[\kappa_c = 6.13 \times 10^{-3}\text{/GPa}\]
\[a= 3.8846 \text{\AA}\]
\[c= 9.5486 \text{\AA}\]
The calculated strain derivatives with different assumptions of Poisson's ratios are shown below.

\centering
\begin{tabular}{cccc}
\\
  \hline
   $\nu_{21},\nu_{31}$& 0.15 & 0.2 & unit\\
   \hline
  s$_{11}$ & 5.69 $\times10^{-3}$   & 6.65 $\times10^{-3}$& /GPa \\
  s$_{12}$ & -0.85$\times10^{-3}$ & -1.33$\times10^{-3}$ & /GPa\\
  s$_{13}$ & -0.85$\times10^{-3}$ & -1.33$\times10^{-3}$ & /GPa\\
  s$_{33}$ & 7.84$\times10^{-3}$ & 8.79$\times10^{-3}$ & /GPa\\
  \hline
  d$T_c$/d$\epsilon_1$ & -490 & -435 & K \\
  d$T_c$/d$\epsilon_3$ & 352 & 278 & K\\
  \hline
  d$T_c$/da & -126 & -111.6 & K/\AA \\
  d$T_c$/dc & 36.8 & 29.1 & K/\AA \\
  \hline
\end{tabular}


\begin{thebibliography}{99}
\bibitem{quench} DAC was first warmed to 180~K and then pressure was immediately released, which was then measured at 90~K to be 1.9~GPa.
\bibitem{Poisson} In fact, the result is insensitive to the choice of Poisson's ratio as long as it is between 0.15-0.3.
\end{thebibliography}
\end{document}